\documentclass{aastex631}

\newcommand{\oiii}{\hbox{[O\,{\sc iii}]}}
\newcommand{\ha}{\hbox{H$\alpha$}}
\newcommand{\hb}{\hbox{H$\beta$}}
\newcommand{\oii}{\hbox{[O\,{\sc ii}]}}
\newcommand{\sii}{\hbox{[S\,{\sc ii}]}}
\newcommand{\nii}{\hbox{[N\,{\sc ii}]}}

\begin{document}

\title{Different influence of gas accretion on the evolution of star-forming and non-star-forming galaxies}

\author[0009-0005-9342-9125]{Min Bao}
\affiliation{School of Physics and Technology, Nanjing Normal University, Nanjing 210023, China}
\affiliation{School of Astronomy and Space Science, Nanjing University, Nanjing 210023, China}

\author{Wenlong Zhao}
\affiliation{School of Physics and Technology, Nanjing Normal University, Nanjing 210023, China}

\author[0000-0002-9244-3938]{Qirong Yuan}
\affiliation{School of Physics and Technology, Nanjing Normal University, Nanjing 210023, China}

\correspondingauthor{Qirong Yuan}
\email{yuanqirong@njnu.edu.cn}

\begin{abstract}

Using integral field spectroscopic data from the Mapping Nearby Galaxies at Apache Point Observatory survey, we investigate the spatially resolved properties and empirical relations of a star-forming galaxy and a non-star-forming galaxy hosting counter-rotating stellar disks (CRDs). The DESI $g, r, z$ color images reveal no evidence of merger remnants in either galaxy, suggesting that gas accretion fuels the formation of CRDs. Based on the visible counter-rotation in the stellar velocity field, we can fit a spatial boundary to distinguish the inner and outer regions dominated by two stellar disks in each galaxy. In the inner region of the star-forming CRDs, stars are co-rotating with ionized gas, and the stellar population is younger. Comparison of the star-forming main sequence relations between the inner and outer regions reveals enhanced star formation in the inner region. Given the abundant pre-existing gas in the star-forming galaxy, collisions between pre-existing and external gas efficiently consume angular momentum, triggering star formation in the inner region. Conversely, in the outer region of the non-star-forming CRDs, stars are co-rotating with ionized gas, and the stellar population is younger. Comparison of the stellar mass-metallicity relations between the inner and outer regions indicates enriched gas-phase metallicity in the outer region. Considering the less abundant pre-existing gas in the non-star-forming galaxy, external gas could preserve angular momentum, fueling star formation in the outer region. Overall, gas accretion exhibits different influence on the evolution of star-forming and non-star-forming galaxies. 

\end{abstract}


\section{Introduction} \label{sec:intro}

Gas accretion, known as a key process in baryon cycle, can provide material for galaxy evolution \citep{2011A&A...535A..72H, 2013ApJ...773....3C, 2014A&ARv..22...71S}. On one hand, metal-poor gas in the circumgalactic medium (CGM) is accreted into galaxies, dilutes the gas-phase metallicity in the interstellar medium (ISM), and fuels the star formation. On the other hand, the metals produced by stars are released into the ISM through processes including stellar winds or supernova explosions. Furthermore, the intensive star formation in the galactic center can trigger outflows in galactic-scale, which entrain the metals from the ISM \citep{2005ARA&A..43..769V}, and generate feedback on star formation in the ISM. The impacts of gas accretion on the star formation and metallicity inside galaxies, as well as its influence on galaxy evolution are still open questions.

Based on single-fiber observations, \cite{2013ApJ...765..140A} discovered a negative correlation between gas-phase metallicity and star formation rate, verifying the contribution of gas accretion to galaxy evolution. With the development of long-slit spectroscopic and integral field spectroscopic instruments, the existence of kinematically misaligned galaxies has been revealed \citep{2014ASPC..486...51C}, which are characterized by the misaligned rotations between two components such as gas-gas, gas-star, or star-star. The misaligned rotations are defined by $\Delta \phi \equiv |\phi_{1} - \phi_{2}| > 30 ^{\circ}$, where $\phi_{1}$ and $\phi_{2}$ represent the position angles of two components. The position angle is defined by the counterclockwise angle between the north and a line bisecting the velocity field on the receding side. The previous studies indicated that these galaxies have undergone gas acquisition events, including mergers and gas accretion, which deliever reversed angular momentum from external. Numerical simulations suggested that gas accretion from the cosmic web dominates the galaxy growth with contribution $\sim$3 to 4 times more than mergers \citep{2014A&ARv..22...71S}. Therefore, the misaligned galaxies in the local Universe can serve as valuable laboratories for investigating the influence of gas accretion on galaxy evolution.

In previous studies, the detection rate of gas-star misalignment in the non-star-forming galaxies has been reported to range from approximately 20\% to 50\% \citep{2006MNRAS.366.1151S, 2011MNRAS.417..882D, 2015A&A...582A..21B}, while in the star-forming galaxies, it is notably lower at around 2\% to 5\% \citep{2016MNRAS.463..913J, 2019MNRAS.483..458B}. \cite{2022MNRAS.511..139B} recently analyzed data from the Mapping Nearby Galaxies at Apache Point Observatory (MaNGA) survey, identifying 64 star-star counter-rotating galaxies, and finding frequencies of less than 5\% for the non-star-forming galaxies and less than 1\% for the star-forming galaxies. The counter-rotation is defined by $\Delta \phi \equiv |\phi_{1} - \phi_{2}| > 150 ^{\circ}$. Given the abundant gas content in the star-forming galaxies, collisions between pre-existing and external gas can efficiently consume the angular momentum. Meanwhile, the non-star-forming galaxies have lower gas content, allowing external gas to retain its angular momentum and to manifest as counter-rotation. These differences in pre-existing gas content and efficiency of angular momentum consumption suggest different formation scenarios for the misalignments (including counter-rotations) in the star-forming and non-star-forming galaxies.

\cite{2016MNRAS.463..913J} studied 66 gas-star misaligned galaxies from the MaNGA survey, and found that the star-forming ones exhibit a positive radial gradient in D$_{n}$4000, whereas the non-star-forming ones show a negative radial gradient. They proposed that the misaligned gas in the star-forming galaxies originates from gas-rich satellites or the cosmic web, where the consumption of angular momentum drives gas inflow, subsequently triggering central star formation. Meanwhile, the non-star-forming misaligned galaxies may form from gas-poor progenitors by accreting misaligned gas. As a series, \cite{2022MNRAS.511.4685X} compared the spatially resolved properties of 456 gas-star misaligned galaxies with control galaxies, and found different trends in star-forming and non-star-forming populations. In star-forming population, the misaligned galaxies have enhanced central star formation compared to their control galaxies, while the difference is small in non-star-forming population. These phenomenons support the different formation scenarios for two populations proposed by \cite{2016MNRAS.463..913J}. Furthermore, \cite{2022MNRAS.515.5081Z} compared the global properties of these 456 gas-star misaligned galaxies with control galaxies, suggesting that the star-forming misaligned galaxies form via gas accretion, while the non-star-forming misaligned galaxies may either form through gas accretion or evolve from star-forming misaligned galaxies.

\cite{2022MNRAS.511..139B} investigated the kinematics of 64 star-star counter-rotating galaxies. They found that the ionized gas in half of these galaxies is co-rotating with one of the stellar disks, which hosts younger stellar population. \cite{2022ApJ...926L..13B} collected 101 star-star counter-rotating galaxies from the MaNGA survey, and categorized them into four types based on the velocities of stars and ionized gas. These counter-rotating galaxies are also classified into star-forming and non-star-forming populations. It turned out that the formation scenarios for star-star counter-rotations are different in two populations, which are governed by the efficiency of angular momentum consumption. In the star-forming galaxies with more abundant pre-existing gas, collisions between pre-existing and external gas efficiently consume the angular momentum. As a result, the inflowing gas can trigger star formation in the center, leading to the co-rotating gas with stars in the inner region. Meanwhile, the consumption of angular momentum is less efficient in the non-star-forming galaxies, so that external gas can fuel star formation on the outskirts, leading to co-rotating gas with stars in the outer region.

In this study, we aim to investigate the influence of gas accretion on galaxy evolution, including star formation and metal enrichment in different regions, by comparing spatially resolved properties and empirical relations. We focus on two galaxies, one star-forming (SDSS~J090852.63+445556.1, named SDSS~J0908+4455) and one non-star-forming (SDSS J093053.94+352623.2, named SDSS~J0930+3526), selected from the sample of \cite{2022ApJ...926L..13B} due to their visible counter-rotations between inner and outer regions in the stellar velocity fields. To make comparisons between these two regions, we fit a spatial boundary in the stellar velocity field of each galaxy. The involved data are presented in Section 2. In Section 3, we present the spatially resolved properties and empirical relations. Finally, we discuss the gas accretion fueling formation of counter-rotations and the influence of gas accretion on star formation and metal enrichment in Section 4. Throughout this paper, we adopt a set of cosmological parameters as follows: $H_{0}$ = 70\,km\,s$^{-1}$, $\Omega_{\rm m}$ = 0.30, $\Omega_{\rm \Lambda}$ = 0.70.

\section{The Data} \label{sec:data}

The MaNGA survey is one of the three core programs in the fourth-generation Sloan Digital Sky Survey (SDSS-IV) \citep{2015ApJ...798....7B, 2016AJ....152...83L}. MaNGA utilized the Baryon Oscillation Spectroscopic Survey (BOSS) spectrographs \citep{2013AJ....146...32S} on the 2.5-m Sloan Foundation Telescope \citep{2006AJ....131.2332G}. This survey conducted IFU observations on a representative sample, which consists of approximately 10,010 unique galaxies with a flat stellar mass distribution in $10^{9}-10^{11}\rm~M_{\odot}$ and a redshift distribution in $0.01-0.15$ \citep{2017AJ....154...28B}. Two dual-channel BOSS spectrographs \citep{2013AJ....146...32S} provide simultaneous wavelength coverage from 3,600 to 10,000\AA. The spectral resolution is $\sim$2,000, which allows measurements of all strong emission line species from {\oii}$\lambda$3726 to {\sii}$\lambda$6731.

We obtain the global properties including redshift ($z$), axial ratio ($q = b/a$) and photometric position angle ($\phi$) from NASA-Sloan Atlas (NSA, \citealt{2011AJ....142...31B}). The global stellar mass and attenuation-corrected star formation rate are obtained by MaNGA Data Reduction Pipeline (DRP, \citealt{2016AJ....152...83L}) and Data Analysis Pipeline (DAP, \citealt{2019AJ....158..231W}). The spatially resolved properties, including emission line flux, stellar (or gas) velocity and velocity dispersion, as well as 4000~\AA~break (D$_{n}$4000) indicating the light weighted stellar population age, are extracted from the MaNGA DAP. The gas kinematics is traced by ionized hydrogen ({\ha}), while all the emission line centers are tied together in the velocity space in the MaNGA DAP. In addition, we acquire the stellar mass ($M_{\star}$) in each spaxel from the data cube of \small PIPE3D \normalsize \citep{2016RMxAA..52...21S}, which analyzed the physical properties of stellar populations of a galaxy using the spectral fitting tool \small FIT3D\normalsize.

We use {\ha}/{\hb} ratio to estimate the extinction corrected intensity of emission lines under the Case B \citep{2001PASP..113.1449C}: 
\begin{equation}
  F_{\lambda } =F_{\lambda ,0}\times {10}^{-0.4k(\lambda )E(B-V)},
\label{eq1:select-QG1}
\end{equation}
where $k(\lambda)$ is the Galactic dust attenuation curve, and color excess $E(B-V)=0.934\times {\rm ln}[(F_{H\alpha} /F_{H\beta})/2.86]$.

The star formation rate (SFR) quantifies the star formation activity within 10$^{6-7}$~years. Following the method from \cite{1998ARA&A..36..189K}, we utilize the {\ha} luminosity in each spaxel to estimate the spatially resolved SFR with a Salpeter IMF \citep{1955ApJ...121..161S}: 
\begin{equation}
  SFR(M_\odot\ yr^{-1})=7.9\times 10^{-42}L_{H\alpha }(1-f_{AGN}),
\label{eq1:select-QG2}
\end{equation}
where we consider a correction factor $f_{\rm AGN}$ to mitigate contamination from active galactic nuclei (AGN), as described in Section \ref{Subsec:spatially resolved properties}.

The gas-phase metallicity 12 + $\log$(O/H) serves as a reliable indicator of the star formation history. We use $\rm R_{23}$ indicator from \cite{2004ApJ...613..898T} to estimate the gas-phase metallicity in each spaxel:
\begin{equation}
  12+log(O/H)=9.185-0.313x-0.264x^{2}-0.321x^{3},
\label{eq1:select-QG3}
\end{equation}
where $\rm x\equiv\log\{({\oii}\lambda \lambda 3726, 3729+{\oiii}\lambda \lambda 4959, 5007)/{\hb}\}$.

SDSS~J0908+4455 and SDSS~J0930+3526 have been identified as galaxies hosting counter-rotating stellar disks (CRDs) by \cite{2022ApJ...926L..13B}, characterized by visible counter-rotations between the inner and outer regions in stellar velocity fields. \cite{2022ApJ...926L..13B} found that the external gas, which is counter-rotating with the pre-existing gas, provides material for the formation of CRDs, while the formation scenarios for the star-forming and non-star-forming galaxies are different. The global $M_{\star}$ and SFR indicate that SDSS~J0908+4455 is a star-forming galaxy located in the blue-cloud, while SDSS~J0930+3526 is a non-star-forming galaxy located in the green-valley, approaching the red-sequence. The comparisons on spatially resolved properties and empirical relations between these two galaxies can help verifying the formation scenarios proposed by \cite{2022ApJ...926L..13B}, and investigating the contributions of different scenarios to galaxy evolution.

\section{Results} \label{sec:results}

\subsection{Spatially resolved properties} \label{Subsec:spatially resolved properties}

\begin{figure*}
     \resizebox{1.\textwidth}{!}{\includegraphics{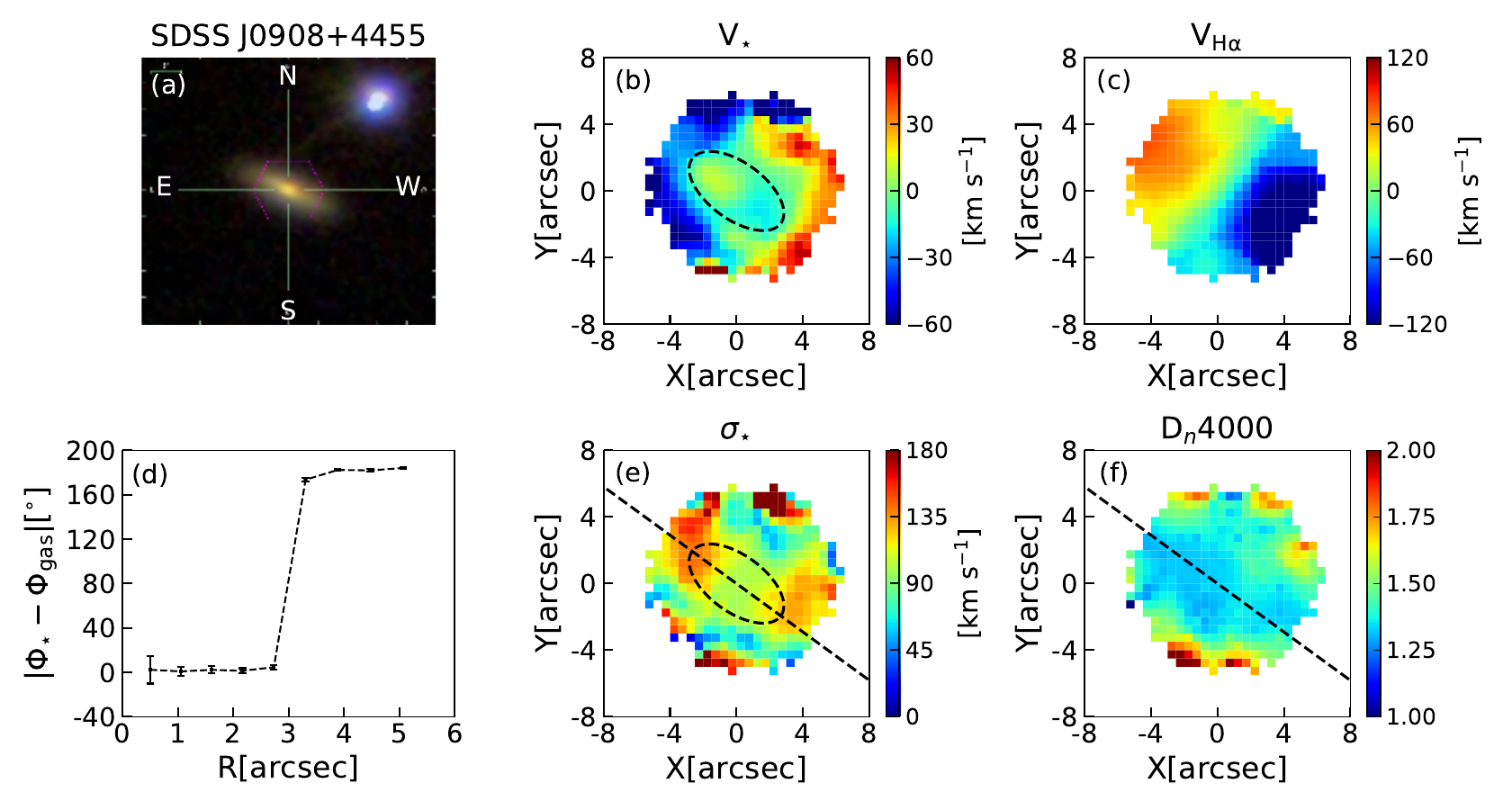}}
    \caption{Spatially resolved properties of SDSS~J0908+4455. (a) SDSS $g, r, i$ color image. (b) The stellar velocity field for spaxels with spectral S/N higher than 3. The black dashed ellipse shows the rotation track where the directions of stellar rotation reverse, and keeps same in panel (e). (c) The gas velocity field for spaxels with {\ha} emission line S/N higher than 3. (d) The differences in position angles between the stars and gas as functions of radii. (e) The stellar velocity dispersion field for spaxels with spectral S/N higher than 3. The black dashed line shows the major axis, and keeps the same in the following maps of SDSS~J0908+4455. (f) The D$_{n}4000$ map for spaxels with spectral S/N higher than 3.}
    \label{KinematicA}
\end{figure*}

Figure \ref{KinematicA}(a) displays the SDSS $g, r, i$ color image of SDSS~J0908+4455. 
Figure \ref{KinematicA}(b) displays the stellar velocity field for the spaxels with spectral signal-to-noise ratio (S/N) higher than 3, and the same criterion is used in the following maps of stellar properties. Stars in the inner region are counter-rotating with that in the outer region, indicating the existence of two CRDs in this galaxy. Given that MaNGA DAP employed the \small PPXF \normalsize single kinematic component fit in the stellar continuum spectra, one of the stellar disks with stronger absorption dominates the stellar velocity. Therefore, the stellar velocity in the inner region is dominated by one stellar disk (named inner disk), while the stellar velocity in the outer region is dominated by another stellar disk (named outer disk).

Figure \ref{KinematicA}(c) displays the gas velocity field for the spaxels with H$\alpha$ emission line S/N higher than 3. Comparing the stellar and gas velocity fields, we find that the inner disk is co-rotating with the gas disk, while the outer disk is counter-rotating with the gas disk. We employ \small KINEMETRY \normalsize package \citep{2011MNRAS.414.2923K} to fit the stellar and gas velocity fields in each 0.5$^{\prime\prime}$ (i.e. pixel size) rotation track, and obtain the position angles. We then compute the differences in position angles between stars and gas, $\Delta\phi = |\phi_{\star} - \phi_{\rm gas}|$, as a function of radii, displayed in Figure \ref{KinematicA}(d). The differences reverse from $\Delta\phi\sim$0$^{\circ}$ to $\Delta\phi\sim$180$^{\circ}$ at a radius of $R\sim3^{\prime\prime}$. We take this radius as a spatial boundary for the regions dominated by inner and outer disks. A rotation track with $R\sim3^{\prime\prime}$ is over-plotted onto the stellar velocity field as a black dashed ellipse in Figure \ref{KinematicA}(b).

Figure \ref{KinematicA}(e) displays the stellar velocity dispersion field. The black dashed line in Figure \ref{KinematicA}(e) represents the photometric major axis (hereafter major axis) of SDSS~J0908+4455, and the black dashed ellipse is the same as Figure \ref{KinematicA}(b). The stellar velocity dispersion is high in two centrosymmetric regions (i.e. 2$\sigma$ regions) along the major axis, coinciding with the boundary between the inner and outer regions. The absorption strengths of two stellar disks are comparable at this boundary, leading to broadened absorption lines and resulting in high velocity dispersions. Figure \ref{KinematicA}(f) displays the D$_{n}$4000 map, which describes the star formation intensity in Gyr-timescale. Along the major axis (black dashed line),  D$_{n}$4000 exhibits a low value of $\sim$1.3 in the galactic disk, including the 2$\sigma$ regions.

\begin{figure*}
     \resizebox{1.\textwidth}{!}{\includegraphics{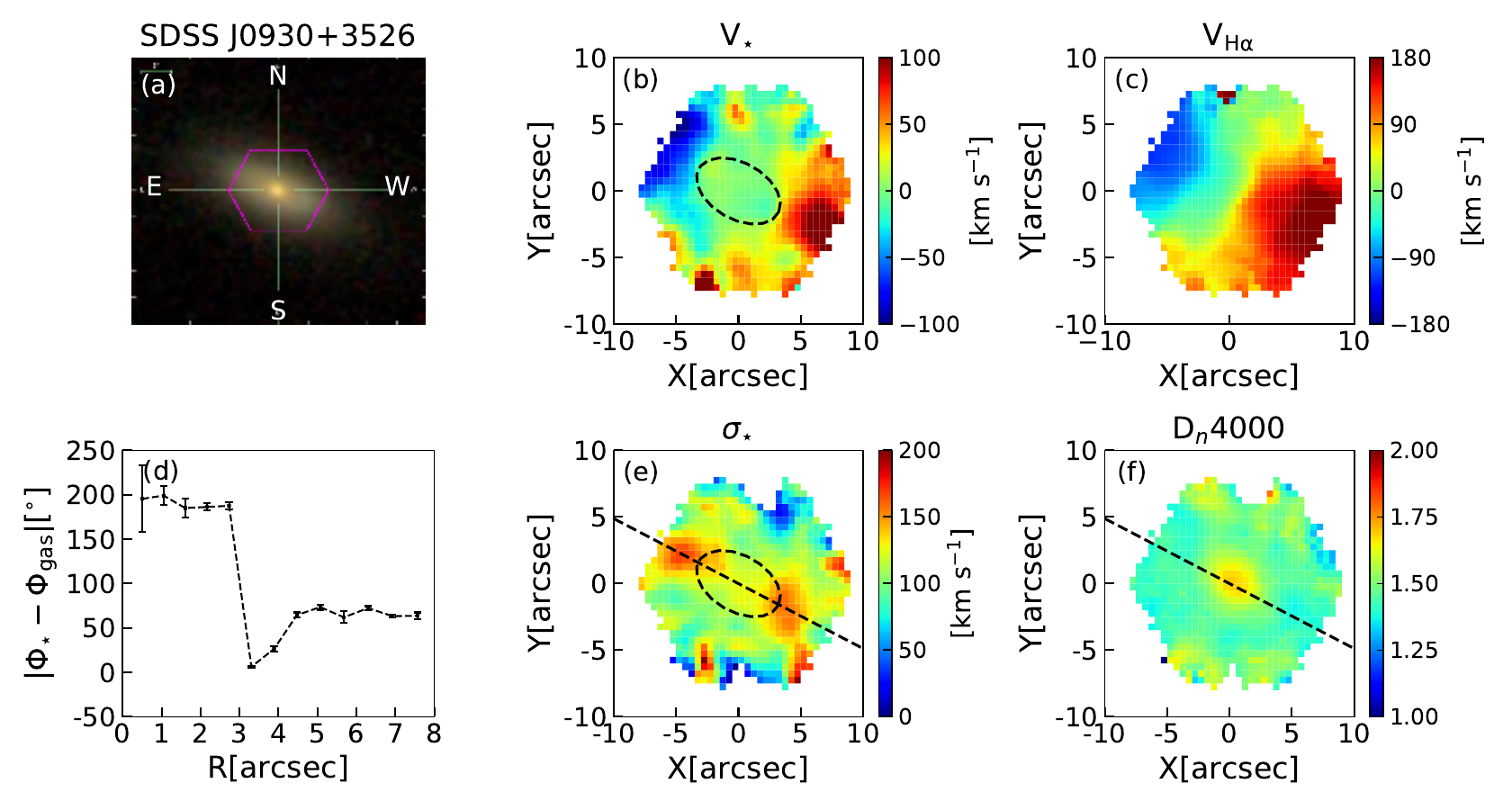}}
    \caption{Spatially resolved properties of SDSS~J0930+3526.  (a) SDSS $g, r, i$ color image. (b) The stellar velocity field for spaxels with spectral S/N higher than 3. The black dashed ellipse shows the rotation track where the directions of stellar rotation reverse, and keeps same in panel (e). (c) The gas velocity field for spaxels with {\ha} emission line S/N higher than 3. (d) The differences in position angles between the stars and gas as functions of radii. (e) The stellar velocity dispersion field for spaxels with spectral S/N higher than 3. The black dashed line shows the major axis, and keeps the same in the following maps of SDSS~J0930+3526. (f) The D$_{n}4000$ map for spaxels with spectral S/N higher than 3.}
    \label{KinematicB}
\end{figure*}

Figure \ref{KinematicB}(a) displays the SDSS $g, r, i$ color image of SDSS~J0930+3526. 
Figures \ref{KinematicB}(b) and \ref{KinematicB}(e) display the stellar velocity and velocity dispersion fields. On one hand, the stellar kinematics is similar to SDSS~J0908+4455 in: (1) counter-rotating inner and outer disks shown in the stellar velocity field (Figure \ref{KinematicB}b); (2) high stellar velocity dispersion at the boundary of two disks (Figure \ref{KinematicB}e). On the other hand, the difference from SDSS~J0908+4455 is that the inner disk is counter-rotating with the gas disk (Figure \ref{KinematicB}c), while the outer disk is co-rotating with the gas disk. We applied the same method to fit the stellar and gas rotations in SDSS~J0930+3526. The differences in position angles between stars and gas in Figure \ref{KinematicB}(d) reverses from $\Delta\phi\sim$180$^{\circ}$ to $\Delta\phi\sim$0$^{\circ}$ at a radius of $R\sim3^{\prime\prime}$. Rotation tracks with $R\sim3^{\prime\prime}$ are over-plotted onto the stellar kinematic fields in Figures \ref{KinematicB}(b) and \ref{KinematicB}(e). Figure \ref{KinematicB}(f) displays the D$_{n}$4000 map. Along the major axis (black dashed line), D$_{n}$4000 presents the highest value of $\sim$1.7 in the central region, while it has a lower value of $\sim$1.4 in the galactic disk, including the 2$\sigma$ regions.

\begin{figure*}
     \centering\resizebox{0.8\textwidth}{!}{\includegraphics{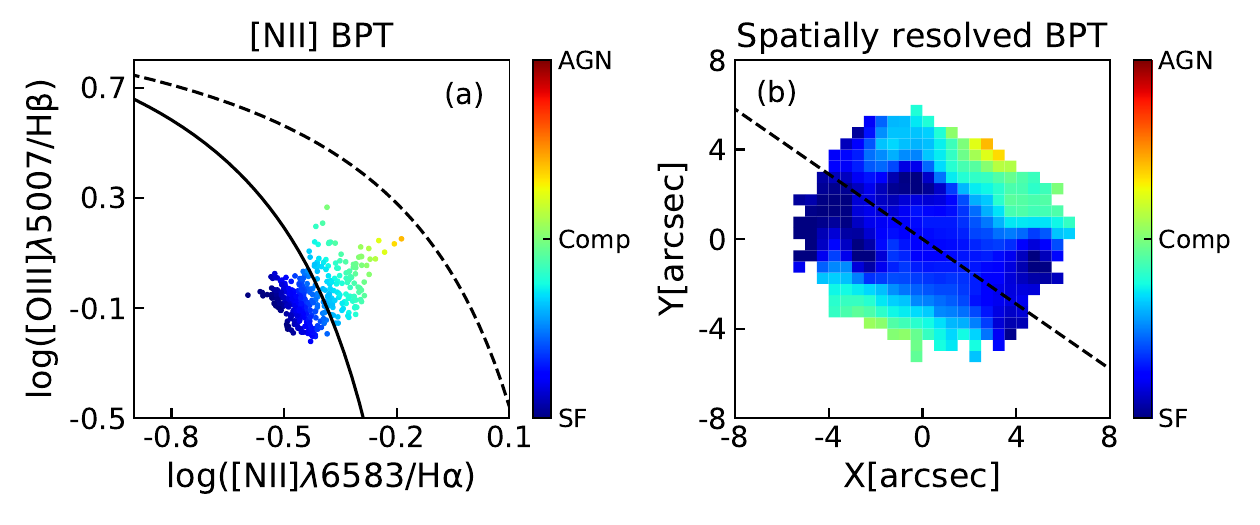}}
     \centering\resizebox{0.8\textwidth}{!}{\includegraphics{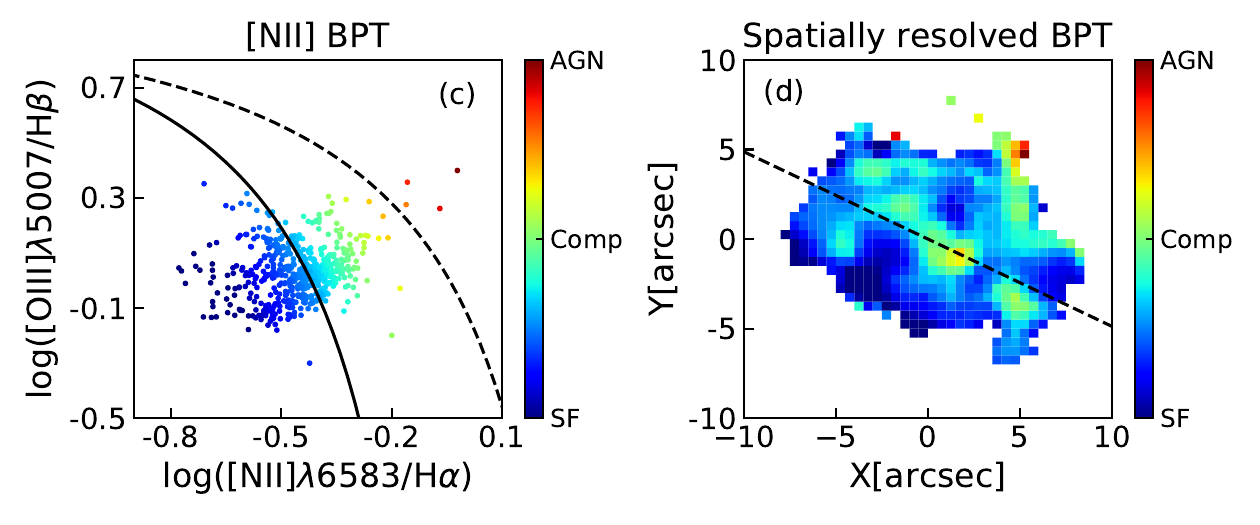}}
    \caption{BPT diagnostic and spatially resolved maps of SDSS~J0908+4455 and SDSS~J0930+3526, for the spaxels with {\hb}, {\oiii}$\lambda$5007, {\ha}, and {\nii}$\lambda$6583 emission line S/Ns higher than 3. (a) The BPT diagnostic map of SDSS~J0908+4455. The black solid curve shows the demarcation between the star-forming and composite regions defined by \cite{2001ApJ...556..121K}, and the black dashed curve shows the demarcation between the composite and AGN regions defined by \cite{2003MNRAS.346.1055K}. The blue and green dots mark the spaxels in star-forming and composite regions that are color-coded by the distance to the black solid curve. (b) The spatially resolved BPT map of J0908. The color-codes are the same as the panel (a). (c) The BPT diagnostic map of SDSS~J0930+3526. The color- and symbol-codes are the same as panel (a), except for the red dots mark the spaxels in AGN region. (d) The spatially resolved BPT map of SDSS~J0930+3526. The color-codes are the same as panel (c).}
    \label{EmissionA}
\end{figure*}

Mapping diagnostic emission line ratios can provide information on the mechanisms of gas ionization. Figure \ref{EmissionA}(a) displays the {\nii} BPT diagnostic map of SDSS~J0908+4455 for spaxels with {\hb}, {\oiii}$\lambda$5007, {\ha} and {\nii}$\lambda$6583 emission line S/Ns higher than 3, and the same criteria are used in the following BPT maps. The black solid curve shows the demarcation between the star-forming and composite regions defined by \cite{2001ApJ...556..121K}, and the black dashed curve shows the demarcation between the composite and AGN regions defined by \cite{2003MNRAS.346.1055K}. Each dot represents the line ratios measured in corresponding spaxel, with color coded by the distance to the black solid curve. Figure \ref{EmissionA}(b) displays the spatially resolved BPT map of SDSS~J0908+4455, with the black dashed line representing the major axis and the color-codes being the same as Figure \ref{EmissionA}(a). The spaxels along the major axis are characterized by star formation, while the spaxels perpendicular to the major axis locate in the composite region. Figures \ref{EmissionA}(c) and \ref{EmissionA}(d) display the {\nii} BPT diagnostic map and spatially resolved BPT map of SDSS~J0930+3526. The black solid and dashed curves in Figure \ref{EmissionA}(c), as well as the color-codes in Figures \ref{EmissionA}(c) and \ref{EmissionA}(d) are the same as Figure \ref{EmissionA}(a). In Figure \ref{EmissionA}(d), the spaxels along the major axis (black dashed line) are basically characterized by star formation, with a few areas belonging to the composite region.

\begin{figure*}
     \centering\resizebox{0.8\textwidth}{!}{\includegraphics{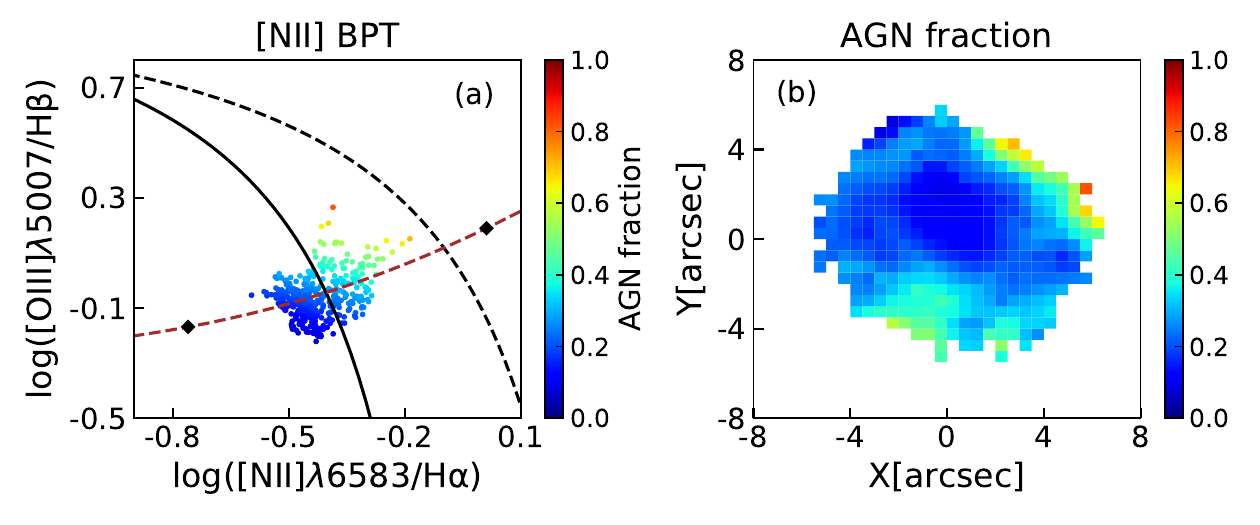}}
     \centering\resizebox{0.8\textwidth}{!}{\includegraphics{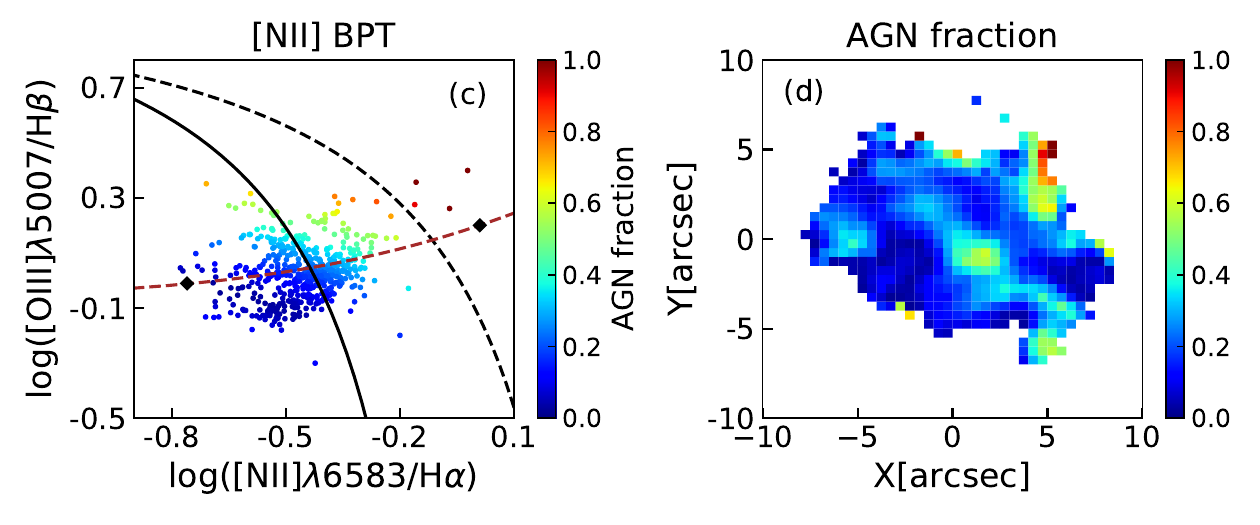}}
    \caption{AGN fractions of SDSS~J0908+4455 and SDSS~J0930+3526, for the spaxels with {\hb}, {\oiii}$\lambda$5007, {\ha}, and {\nii}$\lambda$6583 emission line S/Ns higher than 3. (a) The BPT diagnostic map of SDSS~J0908+4455. The spaxels are color-coded by AGN fraction. The brown dashed curve is the best-fit linear correlation between {\nii}/{\ha} and {\oiii}/{\hb} in the linear space. The black diamond on the left marks the endpoint on the brown dashed curve, where the AGN fraction is set as 0. The black diamond on the right marks the other endpoint, where the AGN fraction is set as 1. (b) The AGN fraction map of SDSS~J0908+4455. The color-codes are the same as panel (a). (c) The BPT diagnostic map of SDSS~J0930+3526. The color- and symbol-codes are the same as panel(a). (d) The AGN fraction map of SDSS~J0908+4455. The color-codes are the same as panel(c).}
    \label{EmissionB}
\end{figure*}

The spatially resolved empirical relations, including star-forming main sequence relation (MSR) and stellar mass-metallicity relation (MZR) can offer insights into the star formation history of a galaxy. However, these relations can be contaminated by AGN. To exclude the contamination, we adopt the methods from \cite{2016MNRAS.462.1616D} and \cite{2019ApJ...881..147S} to quantify the contribution of AGN in each spaxel. Figures \ref{EmissionB}(a) and \ref{EmissionB}(c) display the same {\nii} BPT diagnostic maps as Figures \ref{EmissionA}(a) and \ref{EmissionA}(c). We fit a linear correlation between {\nii}/{\ha} and {\oiii}/{\hb} in the linear space for two galaxies, respectively. The red dashed curves in Figures \ref{EmissionB}(a) and \ref{EmissionB}(c) show the best-fit results for two galaxies. The black diamonds in Figure \ref{EmissionB}(c) mark the endpoints, obtained by projecting the dots with the minimum (P$_{0}$) and maximum (P$_{1}$) {\nii}/{\ha} values onto the curve. Since the distribution of SDSS~J0908+4455 in the {\nii} BPT diagnostic map (Figure \ref{EmissionB}a) is less extended than SDSS~J0930+3526 (Figure \ref{EmissionB}c), we extract the minimum and maximum values from Figure \ref{EmissionB}(c), and project them onto the curve in Figure \ref{EmissionB}(a) (black diamonds). The AGN fraction is defined by $f_{\rm AGN} = L_{\rm P0, N} / L_{\rm P0, P1}$, where L$_{\rm P0, N}$ represents the projected distance along the curve between P$_{0}$ and the Nth dot, and L$_{\rm P0, P1}$ represents the projected distance along the curve between P$_{0}$ and P$_{1}$. The color of each spaxel in Figures \ref{EmissionB}(a) and \ref{EmissionB}(c) is coded by the AGN fraction. Figures \ref{EmissionB}(b) and \ref{EmissionB}(d) display the spatially resolved maps of AGN fraction, where the color-codes are the same as Figures \ref{EmissionB}(a) and \ref{EmissionB}(c).

\subsection{Star-forming main sequence relation \label{subsec:SFR-M}}

\begin{figure*}
     \resizebox{1.\textwidth}{!}{\includegraphics{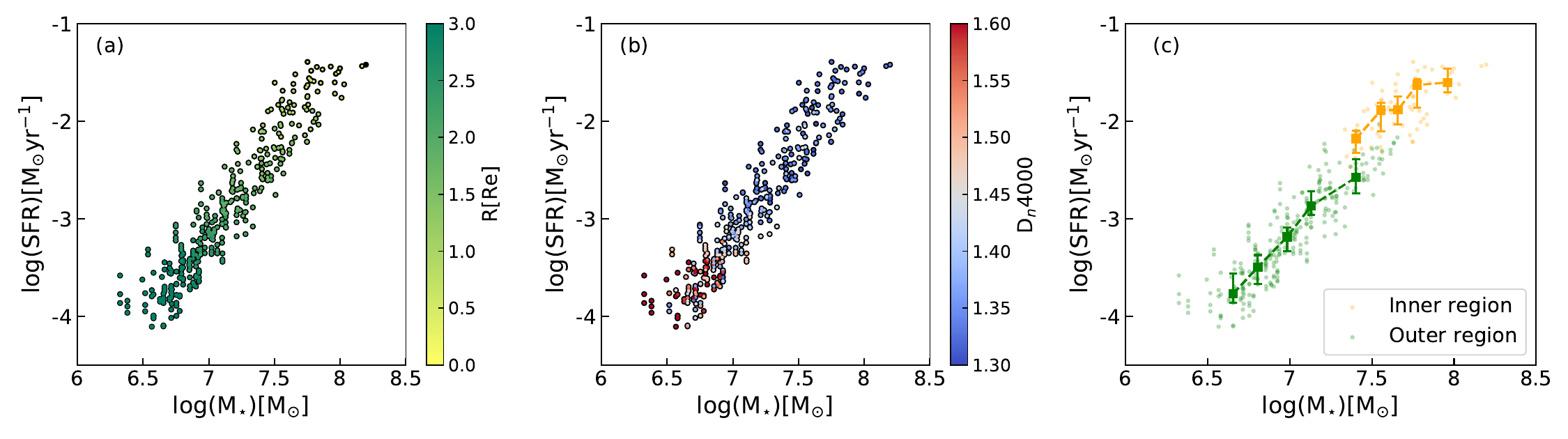}}
    \caption{Star-forming main sequence of SDSS~J0908+4455. (a) The dots show the distributions of spaxels with spectral S/N and {\oii}$\lambda\lambda$3726,3729, {\hb}, {\oiii}$\lambda\lambda$4959,5007, {\ha}, and {\nii}$\lambda$6583 emission line S/Ns higher than 3 in the main sequence plane. The colors are coded by the radii normalized to the effective radius. (b) The distributions are the same as panel (a), while the colors are coded by the D$_{n}4000$ indices. (c) The distributions are the same as panel (a). The light orange dots represent the spaxels that locate in the inner region, while the light green dots represent the spaxels that locate in the outer region. The orange square shows the median $M_{\star}$ and SFR for each $M_{\star}$ bin in the inner region, while the green square marks the median $M_{\star}$ and SFR for each $M_{\star}$ bin in the outer region. The corresponding error bar shows the 25$\%$ to 75$\%$ SFR error in each bin.}
    \label{MSRA}
\end{figure*}

Figure \ref{MSRA} displays the MSR of the spaxels in SDSS~J0908+4455, with spectral S/N and {\oii}$\lambda\lambda$3726,3729, {\hb}, {\oiii}$\lambda\lambda$4959,5007, {\ha}, and {\nii}$\lambda$6583 emission line S/Ns higher than 3, and the same criteria are used in the following MSRs and MZRs. The SFR in each spaxel is estimated following Equation \ref{eq1:select-QG2}, which excludes the contamination from AGN. In Figure \ref{MSRA}(a), each dot represents the location of corresponding spaxel in the main sequence plane, color-coded by radius. The spaxels in SDSS~J0908+4455 follow a tight MSR, characterized by a linear correlation between $M_{\star}$ and SFR. Both $M_{\star}$ and SFR are highest in the center, and monotonously decrease with increasing radius. The dots in Figure \ref{MSRA}(b) share the same positions with Figure \ref{MSRA}(a), while are color-coded by the light weighted stellar population age (D$_{n}$4000). It turns out that the young stellar population locates in the inner region with high $M_{\star}$ and SFR, indicating recent star formation in this region.

In Figure \ref{MSRA}(c), we take the spatial boundary (black dashed ellipse) from Figure \ref{KinematicA}(b) to distinguish the spaxels dominated by the inner and outer disks. The light-orange dots represent the spaxels in the inner region dominated by the inner disk, and the light-green dots represent the spaxels in the outer region dominated by the outer disk. We equally divide the spaxels in the inner or outer regions into five $M_{\star}$ bins. For each bin in the inner region, the orange square shows the median SFR, with the orange bar showing the 25\% to 75\% SFR error. Similarly, the green square shows the median SFR for each bin in the outer region, with the green bar showing the SFR error. Overall, the $M_{\star}$ and SFR in the inner region (light-orange dots) are higher than that in the outer region (light-green dots). Furthermore, the SFR of the bin with $M_{\star}\sim10^{7.5}\rm M_{\odot}$ in the inner region (leftmost orange square) exceeds that in the outer region. Given that the spaxels in the inner and outer regions with $M_{\star}\sim10^{7.5}\rm M_{\odot}$ have continuous radii, presented as unified colors in this bin in Figure \ref{MSRA}(a). The difference in SFR between the inner and outer regions indicates a gap in the star formation intensity between the inner and outer disks, implying the enhanced star formation in the inner disk.

\begin{figure*}
     \resizebox{1.\textwidth}{!}{\includegraphics{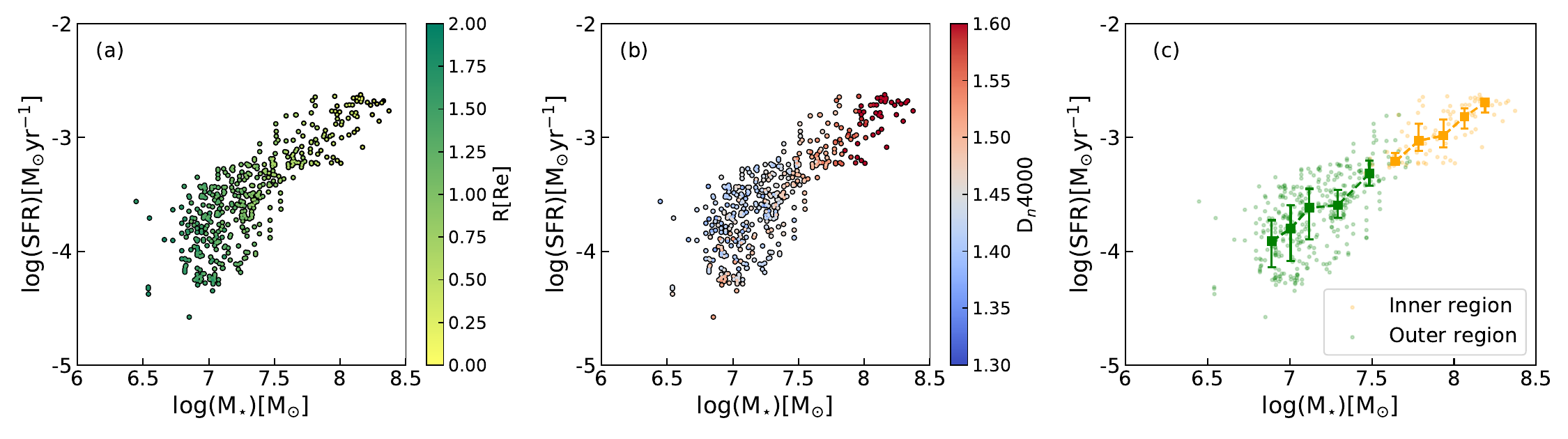}}
    \caption{Star-forming main sequence of SDSS~J0930+3526. (a) The dots show the distributions of spaxels with spectral S/N and {\oii}$\lambda\lambda$3726,3729, {\hb}, {\oiii}$\lambda\lambda$4959,5007, {\ha}, and {\nii}$\lambda$6583 emission line S/Ns higher than 3 in the main sequence plane. The colors are coded by the radii normalized to the effective radius. (b) The distributions are the same as panel (a), while the colors are coded by the D$_{n}4000$ indices. (c) The distributions are the same as panel (a). The light orange dots represent the spaxels that locate in the inner region, while the light green dots represent the spaxels that locate in the outer region. The orange square shows the median $M_{\star}$ and SFR for each $M_{\star}$ bin in the inner region, while the green square marks the median $M_{\star}$ and SFR for each $M_{\star}$ bin in the outer region. The corresponding error bar shows the 25$\%$ to 75$\%$ SFR error in each bin.}
    \label{MSRB}
\end{figure*}

Figure \ref{MSRB} displays the MSR of the spaxels in SDSS~J0930+3526, with the SFR estimated following Equation \ref{eq1:select-QG2}. In Figures \ref{MSRB}(a) and \ref{MSRB}(b), the dots represent the locations of spaxels in the main sequence plane, color-coded by radii and D$_{n}$4000, respectively. Similar to SDSS~J0908+4455 (Figure \ref{MSRA}a), the $M_{\star}$ and SFR decrease with increasing radius in Figure \ref{MSRB}(a). However, the young stellar population locates in the outer region with low $M_{\star}$ and SFR, suggesting that additional star formation could occur recently in this region. The S$\acute{\rm e}$rsic index of SDSS~J0930+3526 equals $\sim$2.41, indicating the existence of a bulge component in this galaxy, which can also be found in the center of SDSS $g, r, i$ color image in Figure \ref{KinematicB}(a). The old stellar population in the bulge dominates the age in the center. In Figure \ref{MSRB}(c), the light-orange and light-green dots representing spaxels in the inner and outer regions, are separated by the spatial boundary (black dashed ellipse) from Figure \ref{KinematicB}(b), and are equally divided into five $M_{\star}$ bins. The orange or green square shows the median SFR in each bin, and implies that the $M_{\star}$ and SFR of inner and outer regions follow a consistent MSR.

\subsection{Stellar mass-metallicity relation \label{subsec:Z-M}}

\begin{figure*}
     \resizebox{1\textwidth}{!}{\includegraphics{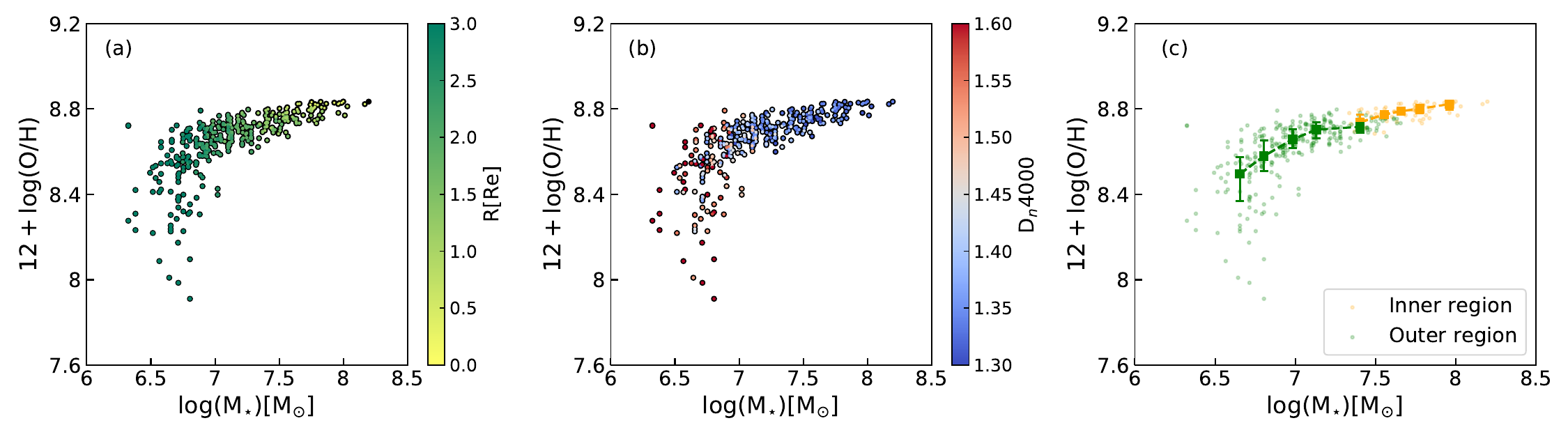}}
    \caption{Stellar mass-metallicity relation of SDSS~J0908+4455. (a) The dots show the distributions of spaxels with spectral S/N and {\oii}$\lambda\lambda$3726,3729, {\hb}, {\oiii}$\lambda\lambda$4959,5007, {\ha}, and {\nii}$\lambda$6583 emission line S/Ns higher than 3 in the stellar mass-metallicity plane. The colors are coded by the radii normalized to the effective radius. (b) The distributions are the same as panel (a), while the colors are coded by the D$_{n}4000$ indices. (c) The distributions are the same as panel (a). The light orange dots represent the spaxels that locate in the inner region, while the light green dots represent the spaxels that locate in the outer region. The orange square shows the median $M_{\star}$ and $12 + \log(\rm O/H)$ for each $M_{\star}$ bin in the inner region, while the green square marks the median $M_{\star}$ and $12 + \log(\rm O/H)$ for each $M_{\star}$ bin in the outer region. The corresponding error bar shows the 25$\%$ to 75$\%$ $12 + \log(\rm O/H)$ error in each bin.}
    \label{MZRA}
\end{figure*}

Figure \ref{MZRA} displays the MZR of spaxels in SDSS~J0908+4455. The gas-phase metallicity is estimated following Equation \ref{eq1:select-QG3}. In Figures \ref{MZRA}(a) and \ref{MZRA}(b), the colors of dots represent radii and D$_{n}$4000, respectively.  A tight correlation between $M_{\star}$ and metallicity is observed for spaxels with $M_{\star} > 10^{7}~\rm M_{\odot}$, while it becomes scattered for spaxels with $M_{\star} < 10^{7}~\rm M_{\odot}$. The $M_{\star}$ and metallicity are highest in the center, and decrease with increasing radius. Given the young stellar population in the inner region (Figure \ref{MZRA}b), the gas-phase metallicity can be enriched by the recent star formation in this region. Figure \ref{MZRA}(c) displays the MZRs in the inner and outer regions, represented by light-orange and light-green dots. Following the $M_{\star}$ bins in Figures \ref{MSRA}(c), the orange and green squares show the median metallicity in each $M_{\star}$ bin with 25\% - 75\% metallicity error. Different from the gap observed in SFR (Figure \ref{MSRA}c), the gas-phase metallicity of the inner and outer regions are consistent at the boundary, i.e. the inner and outer regions of SDSS~J0908+4455 follow a consistent MZR.

\begin{figure*}
     \resizebox{1\textwidth}{!}{\includegraphics{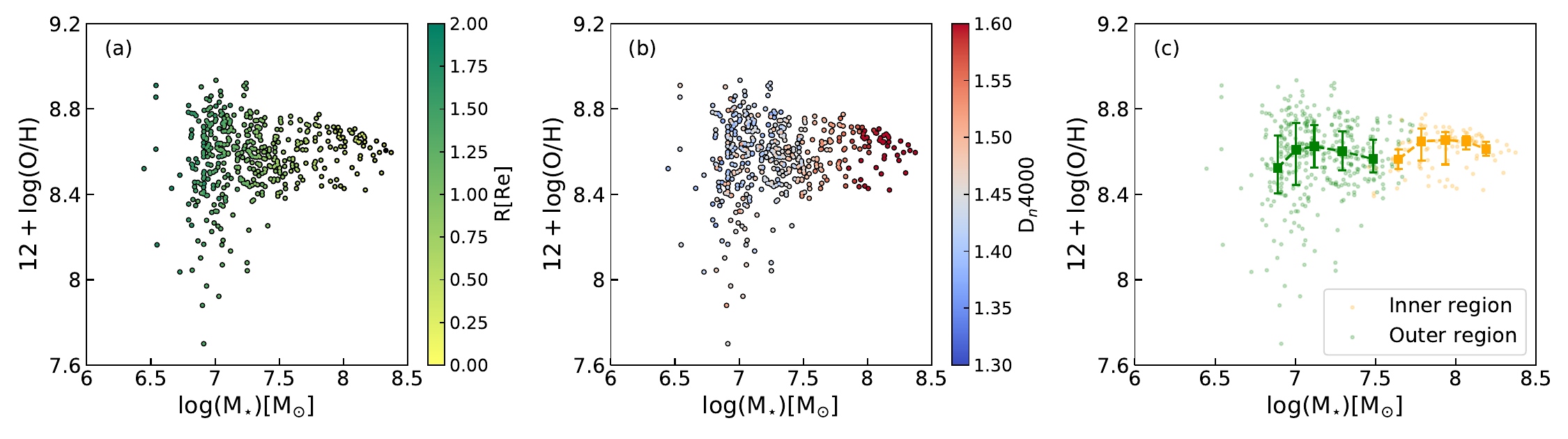}}
     \caption{Stellar mass-metallicity relation of SDSS~J0930+3526. (a) The dots show the distributions of spaxels with spectral S/N and {\oii}$\lambda\lambda$3726,3729, {\hb}, {\oiii}$\lambda\lambda$4959,5007, {\ha}, and {\nii}$\lambda$6583 emission line S/Ns higher than 3 in the stellar mass-metallicity plane. The colors are coded by the radii normalized to the effective radius. (b) The distributions are the same as panel (a), while the colors are coded by the D$_{n}4000$ indices. (c) The distributions are the same as panel (a). The light orange dots represent the spaxels that locate in the inner region, while the light green dots represent the spaxels that locate in the outer region. The orange square shows the median $M_{\star}$ and $12 + \log(\rm O/H)$ for each $M_{\star}$ bin in the inner region, while the green square marks the median $M_{\star}$ and $12 + \log(\rm O/H)$ for each $M_{\star}$ bin in the outer region. The corresponding error bar shows the 25$\%$ to 75$\%$ $12 + \log(\rm O/H)$ error in each bin.}
    \label{MZRB}
\end{figure*}

Figure \ref{MZRB} displays the MZR of spaxels in SDSS~J0930+3526. The gas-phase metallicity is estimated following Equation \ref{eq1:select-QG3}. In Figures \ref{MZRB}(a) and \ref{MZRB}(b), the colors of dots represent radii and D$_{n}$4000. In Figure \ref{MZRB}(c), the $M_{\star}$ bins and color-codes are the same as Figures \ref{MSRB}(c). The $M_{\star}$ of inner region is higher than that of outer region in Figure \ref{MZRB}(c), while the gas-phase metallicity of two regions are comparable. Previous studies have demonstrated positive correlations between $M_{\star}$ and metallicity both in global and spatially resolved analyses \citep{2004ApJ...613..898T, 2008ApJ...681.1183K, 2016MNRAS.463.2513B, 2024MNRAS.529.4993L}. Combining the young stellar population in the outer region (Figure \ref{MZRB}b) of SDSS~J0930+3526, the additional star formation in this region could also contribute to the metal enrichment.

\section{Discussion} \label{sec:discussion}

The evolution of a galaxy, known as a complex system, is regulated by both internal and external processes. External processes, including galaxy mergers and gas accretion, can deliever gas with misaligned angular momentum. Once the external gas becomes more abundant than the pre-existing gas in progenitor, it can fuel the formation of misaligned structures, including counter-rotations. In observations, agreement still hasn't been reached on whether mergers or gas accretion dominate the formation of CRDs. Previous studies have proposed the formation scenarios for CRDs as major merger \citep{2009MNRAS.393.1255C} or minor merger \citep{2014A&A...570A..79P, 2014MNRAS.441.2212F, 2024ApJ...962...27K}, gas accretion from satellite \citep{2011MNRAS.412L.113C, 2013A&A...549A...3C} or enviroment \citep{2013A&A...549A...3C, 2014A&A...570A..79P, 2014MNRAS.441.2212F, 2018A&A...616A..22P}, as well as merger with consequent slow gas accretion \citep{2016MNRAS.461.2068K}.  

\begin{figure*}
     \centering\resizebox{0.9\textwidth}{!}{\includegraphics{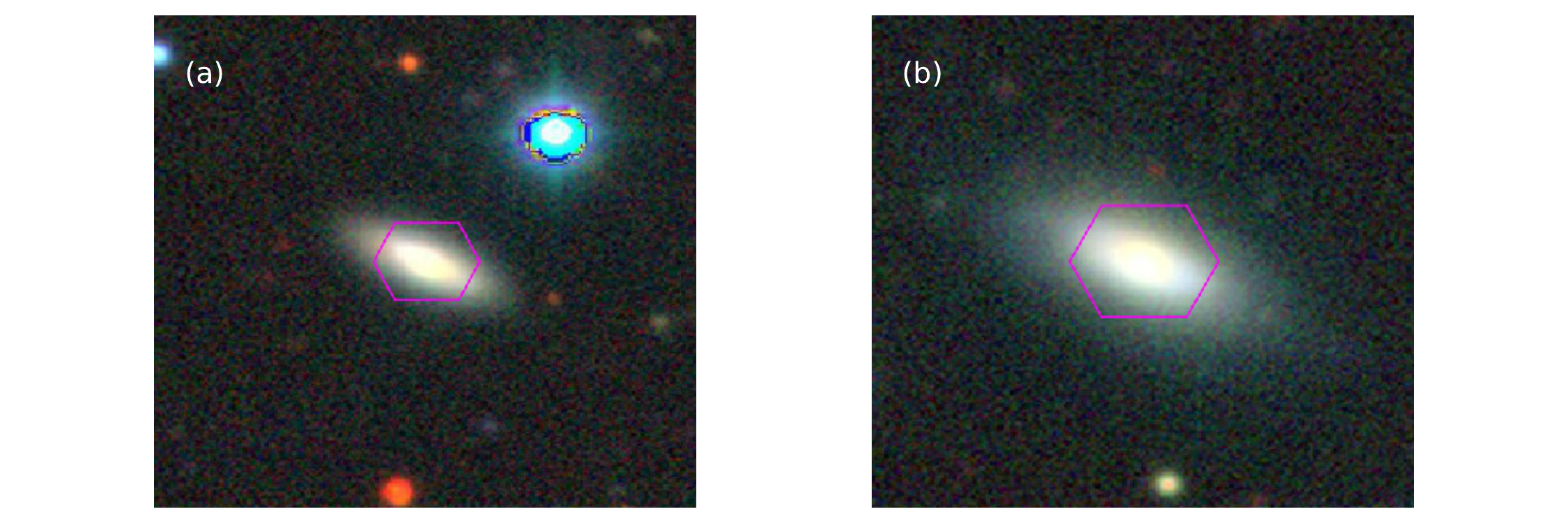}}
    \caption{DESI $g, r, z$ color images, with the purple hexagon displaying the field-of-view of MaNGA survey. (a) The image of SDSS~J0908+4455. (b) The image of SDSS~J0930+3526.}
    \label{DESI}
\end{figure*}

\cite{2014A&A...566A..97J} simulated mergers with different mass ratios, incorporating physical processes such as gas cooling, star formation, and supernova feedback, and investigated the corresponding merger times. For a surface brightness limit of 25~mag~arcsec$^{-2}$, the merger remnant features can be observed in $\sim$2.57~Gyr for equal-mass mergers, while can be observed in $\sim$5.19~Gyr for unequal-mass mergers. Figures \ref{DESI}(a) and \ref{DESI}(b) display the DESI images of SDSS~J0908+4455 and SDSS~J0930+3526, which are much deeper than 25~mag~arcsec$^{-2}$ in $r$-band. Equal-mass mergers, known as disruptive events, can destroy the present morphology of galaxies and tend to form elliptical galaxies \citep{2005A&A...437...69B, 2011A&A...530A..10Q}. However, disk structures are present in both DESI images in Figures \ref{DESI}(a) and \ref{DESI}(b), which exclude the possiblity of CRDs contributed by equal-mass mergers. Following the method from \cite{2021MNRAS.501...14L}, we collect the $g$- and $r$-band images, convolve one band having smaller seeing with a Gaussian function to match the point spread function of the other band, and stack two images to improve S/N. By visual inspection, we find no merger remnant features in these two galaxies, indicating that SDSS~J0908+4455 and SDSS~J0930+3526 haven't experienced merger events at least within the past 5~Gyr.

\cite{2016NatCo...713269C} investigated the gas-star misalignment in 9 star-forming galaxies, and found that the stellar population is younger and star formation is more active in the center compared to the outskirts. \cite{2016MNRAS.463..913J} enlarged the sample to 10 star-forming, 26 green-valley, and 30 quiescent gas-star misaligned galaxies. They also observed a positive radial gradient of D$_{n}$4000 in the star-forming misaligned galaxies, while the D$_{n}$4000 follows a negative radial gradient in the non-star-forming ones. \cite{2022MNRAS.511.4685X} further enlarged the sample to 72 star-forming, 142 green-valley and 242 quiescent gas-star misaligned galaxies. They found enhanced star formation in the center of the star-forming misaligned galaxies, manifested as positive a radial gradient of D$_{n}$4000 and a negative radial gradient of star formation rate surface density. All these phenomenons support a scenario that accreted gas with misaligned angular momentum is more abundant than the pre-existing gas in the progenitors, resulting in kinematic misalignment between gas and stars. Additionally, the pre-existing gas in the star-forming galaxies is abundant, so that gas collisions can efficiently consume angular momentum, trigger gas inflow and star formation in the center.

\begin{figure*}
     \centering\resizebox{0.9\textwidth}{!}{\includegraphics{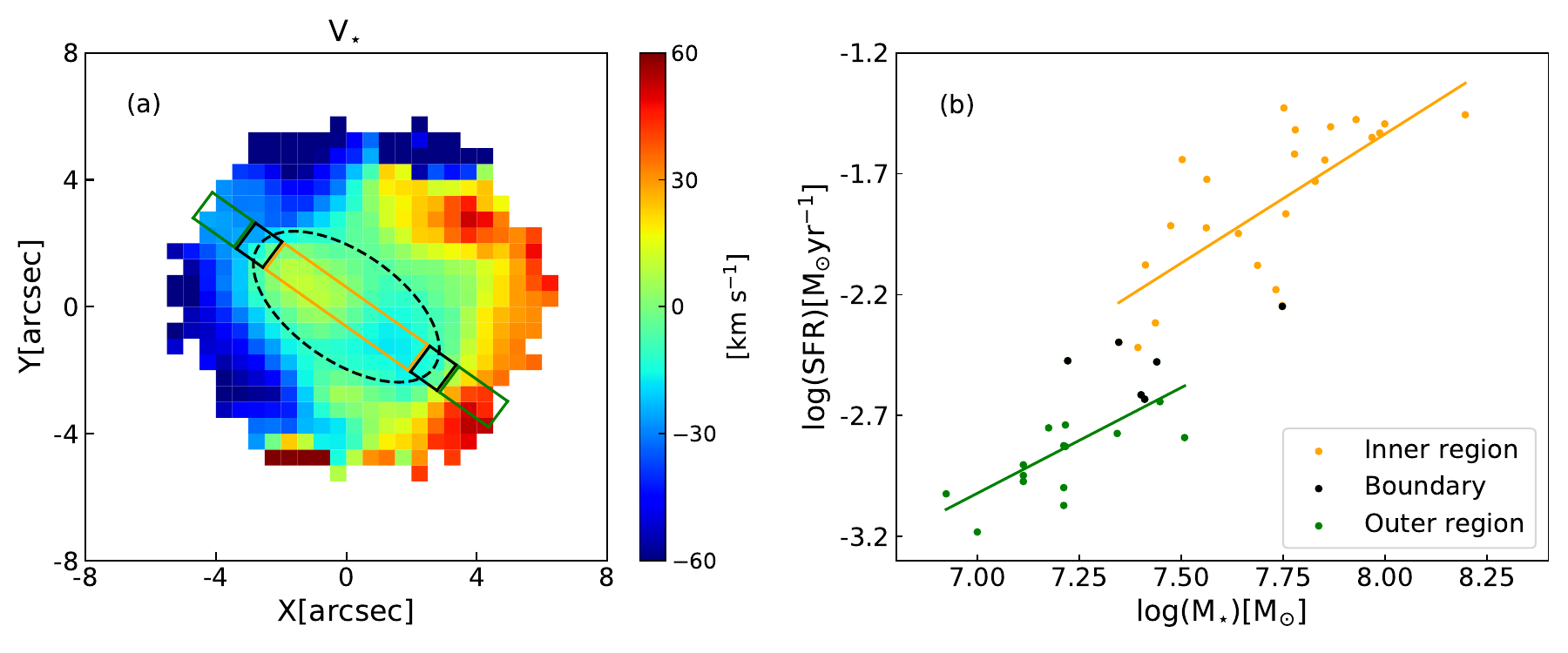}}
    \caption{The star-forming main sequence relation along the major axis of SDSS~J0908+4455. (a) The stellar velocity field of SDSS~J0908+4455 with spectral S/N higher than 3. The black dashed ellipse is the boundary same as Figure \ref{KinematicA}(b). The rectangles are centered on the major axis and extend $\pm$0.5~arcsec along the direction of minor axis, with orange, green, black colors corresponding to the inner region, outer region, boundary between them. (b) The star-forming main sequence relation inside the rectangles with spectral S/N and {\oii}$\lambda\lambda$3726,3729, {\hb}, {\oiii}$\lambda\lambda$4959,5007, {\ha}, and {\nii}$\lambda$6583 emission line S/Ns higher than 3. The green, black, orange dots show the distributions of the spaxels in rectangles with corresponding colors. The green and orange solid lines are the best-fit linear correlations for the green and orange dots.}
    \label{MSRMajorA}
\end{figure*}

The younger stellar population (Figure \ref{MSRA}b) and enhanced star formation (Figure \ref{MSRA}c) in the inner region of SDSS~J0908+4455 support this scenario. To make a clear comparison between the inner and outer regions, we exclude spaxels at the boundary in the main sequence plane. Figure \ref{MSRMajorA} displays the stellar velocity field of SDSS~J0908+4455. The black dashed ellipse represents the spatial boundary, the same as Figure \ref{KinematicA}(b). Rectangles are centered on the major axis and extends $\pm$0.5~arcsec along the direction of minor axis. The orange and green rectangles mark the areas dominated by the inner and outer disks, while the black rectangles mark the boundary areas. Figure \ref{MSRMajorA}(b) displays the MSRs in different areas. Dots represent spaxels extracted from different rectangles in Figure \ref{MSRMajorA}(a), with the yellow, green and black colors corresponding to inner region, outer region and boundary. The orange and green lines show the best-fit linear correlations for the inner and outer regions, which have comparable slopes equaling 1.07 and 0.87, respectively. Meanwhile, an obvious gap of $\sim$0.5~dex exists in SFR between two regions in Figure \ref{MSRMajorA}(b), quantitatively demonstrating the star formation enhancement in the inner disk.

\begin{figure*}
     \centering\resizebox{0.6\textwidth}{!}{\includegraphics{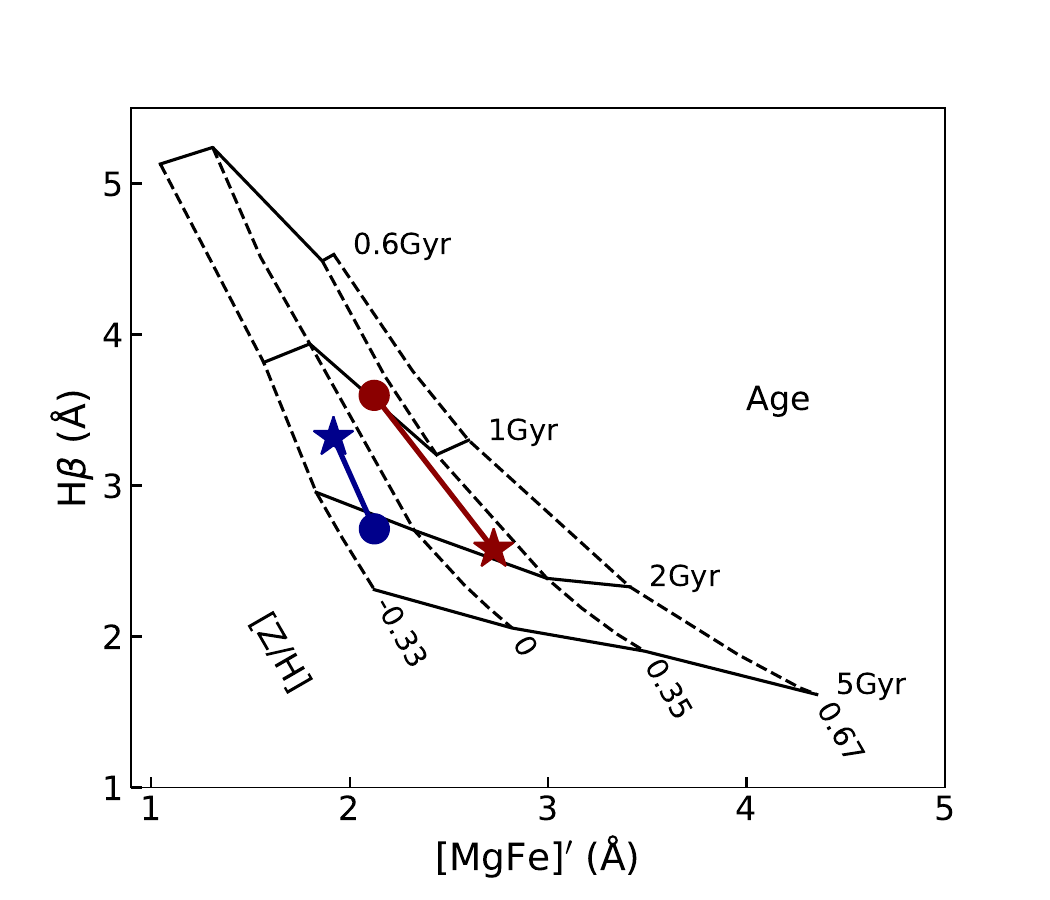}}
    \caption{The stellar properties of the inner and outer regions of SDSS~J0908+4455 and SDSS~J0930+3526. The marks show the {\hb} indices as functions of [MgFe]$^{\prime}$ indices, with the model-grid showing the prediction from single stellar population models \citep{2011MNRAS.412.2183T}. The blue star and circle show the average indices of the inner and outer regions of SDSS~J0908+4455, and the red star and circle show the average indices of the inner and outer regions of SDSS~J0930+3526.}
    \label{stellar_properties}
\end{figure*}

Gas-phase metallicity can provide insights into the origin of external gas. \cite{2017MNRAS.469..151B} studied the gas-phase metallicity in star-forming regions across a representative sample of 550 MaNGA galaxies with $M_{\star} \sim 10^{9}-10^{11.5}~\rm M_{\odot}$. They found the radial metallicity gradient lying flat at $M_{\star} \sim 10^{9}~\rm M_{\odot}$, while exhibiting slope as -0.14~dex~$Re^{-1}$ at $M_{\star} \sim 10^{10.5}~\rm M_{\odot}$. The metallicity gradient approximately equals -0.05~dex~$Re^{-1}$ at $M_{\star} \sim 10^{9.5}-10^{9.75}~\rm M_{\odot}$ \citep{2017MNRAS.469..151B}. For SDSS~J0908+4455, the MaNGA DRP estimates its global stellar mass to be $M_{\star} \sim 10^{9.6}~\rm M_{\odot}$. Additionally, the effective radius of this galaxy is $\sim$3.0~arcsec, coinciding with the radius of boundary between the inner and outer regions. In other words, the inner disk happen to dominate the region inside the effective radius. The MZR of inner region in Figure \ref{MZRA}(c) implies a radial metallicity gradient of $\sim$-0.1~dex~$Re^{-1}$, steeper than the gradient of galaxies with similar stellar mass, but comparable to that of more massive galaxies \citep{2017MNRAS.469..151B}. Gas accretion from the cosmic web can deliever metal-poor gas into SDSS~J0908+4455, which dilutes the gas-phase metallicity in the disk. Meanwhile, the enhanced star formation in the inner disk can enrich the gas-phase metallicity in the center, which steepens the radial metallicity gradient.

To estimate the stellar properties of the inner and outer regions of two galaxies, we collect the stellar indices from the MaNGA DRP, and calculate the combined magnesium-iron index [MgFe]$^{\prime}$ as $\sqrt{\rm Mgb \cdot (0.72 \cdot Fe5270 + 0.28 \cdot Fe5335)}$, which is independent of $\alpha$-enhancement and can be an effective indicator of the stellar metallicity  \citep{2003A&A...401..429T}. Figure \ref{stellar_properties} displays the {\hb} indices as functions of [MgFe]$^{\prime}$ indices, with the model-grid showing the prediction from single stellar population models \citep{2011MNRAS.412.2183T}. The blue star and circle show the average indices of the inner and outer regions of SDSS~J0908+4455, and the red star and circle show the average indices of the inner and outer regions of SDSS~J0930+3526. In SDSS~J0908+4455, comparing the marks with the model-grid, the stellar population ages of the inner and outer regions equal $\sim$1.53~Gyr and $\sim$2.64~Gyr, while the stellar metallicity (i.e. [Z/H]) of the inner and outer regions are comparable, equaling $\sim$-0.14 and $\sim$-0.19. The younger stellar population in the inner region supports the scenario that the accreted gas triggers star formation in the inner region. In SDSS~J0930+3526, by similar comparisons, the stellar population ages of the inner and outer regions equal $\sim$1.90~Gyr and $\sim$0.98~Gyr, while the stellar metallicity of the inner and outer regions are comparable, equaling $\sim$0.24 and $\sim$0.19. Conversely, the younger stellar population in the outer region supports the scenario that the accreted gas triggered star formation in the outer region.

In Figure \ref{MSRB}(c), the MSRs of inner region (orange squares) and outer region (green squares) in SDSS~J0930+3526 follow a consistent gradient. The pearson coefficient ($r$), ranging from 0 to 1, is a parameter quantifying the relativity between two parameters. A value closer to 1 indicates a strong linear correlation between the two parameters. The r value between $M_{\star}$ and SFR (orange and green dots) in Figure \ref{MSRB}(c) equals $\sim$0.93, indicating a tight linear correlation. Thus, the inner and outer regions in SDSS~J0930+3526 follow an unified MSR. In Figure \ref{MZRB}(c), $M_{\star}$ of the inner region (green dots) is generally higher than that of the outer region (orange dots). However, the gas-phase metallicity of two regions turns out to be comparable. The r value between $M_{\star}$ and metallicity equals $\sim$0.08, indicating no linear correlation when considering the inner and outer regions as a whole. Combining the younger stellar population in the outer region in Figure \ref{stellar_properties}, we propose that the external gas fueled the additional star formation in the outer region of SDSS~J0930+3526 and contributed to metal enrichment.

\section{Summary} \label{sec:summary}

In this paper, we studied the spatially resolved physical properties and empirical relations of two galaxies hosting CRDs, one star-forming galaxy and one non-star-forming galaxy. The observational evidences support the formation scenarios for CRDs proposed by \cite{2022ApJ...926L..13B}, and shed light on the internal evolution processes along with the formation of CRDs.

SDSS~J0908+4455 is a star-forming galaxy with more abundant pre-existing gas. Efficient collisions between pre-existing and external gas can consume angular momentum and trigger star formation in the inner region. The observational evidences are as follows: (1) the stellar population in the inner region is younger than that in the outer region; (2) the inner disk is co-rotating with the gas disk; (3) the MSR of inner region has a comparable slope but $\sim$0.5~dex higher SFR compared to that of outer region, which indicates enhanced star formation in the inner region; (4) the MZR of inner region implies a steep radial gradient, which supports gas accretion from the cosmic web plus enhanced star formation in the center.

SDSS~J0930+3526 is a non-star-forming galaxy with less abundant pre-existing gas. The angular momentum of external gas could be reserved. Consequently, the external gas fueled star formation in the outer region, which enriched the gas-phase metallicity in this region. The observational evidences are as follows: (1) the stellar population in the outer region is younger than that in the inner region; (2) the outer disk is co-rotating with the gas disk; (3) the stellar mass of the inner region is generally higher than that of the outer region, while the gas-phase metallicity of these two regions are comparable, which indicates enriched gas-phase metallicity in the outer region.

\begin{acknowledgments}
     \textbf{Acknowledgements:} 
     MB acknowledges support from the National Natural Science Foundation of China, NSFC Grant No. 12303009. QY acknowledges support from the National Natural Science Foundation of China, NSFC Grant Nos. 12273013, 12173020.

     Funding for the Sloan Digital Sky Survey IV has been provided by the Alfred P. Sloan Foundation, the U.S. Department of Energy Office of Science, and the Participating Institutions. SDSS- IV acknowledges support and resources from the Center for High-Performance Computing at the University of Utah. The SDSS web site is www.sdss.org. SDSS-IV is managed by the Astrophysical Research Consortium for the Participating Institutions of the SDSS Collaboration including the
     Brazilian Participation Group, the Carnegie Institution for Science, Carnegie Mellon University, the Chilean Participation Group, the French Participation Group, Harvard-Smithsonian Center for Astrophysics, Instituto de Astrof\'{i}sica de Canarias, The Johns Hopkins University, Kavli Institute for the Physics and Mathematics of the Universe (IPMU) / University of Tokyo, Lawrence Berkeley National Laboratory, Leibniz Institut  f\"{u}r Astrophysik Potsdam (AIP), Max-Planck-Institut  f\"{u}r   Astronomie  (MPIA  Heidelberg), Max-Planck-Institut   f\"{u}r   Astrophysik  (MPA   Garching),
     Max-Planck-Institut f\"{u}r Extraterrestrische Physik (MPE), National Astronomical Observatory of China, New Mexico State University, New York University, University of Notre Dame, Observat\'{o}rio Nacional / MCTI, The Ohio State University, Pennsylvania State University, Shanghai Astronomical Observatory, United Kingdom Participation Group, Universidad Nacional  Aut\'{o}noma de M\'{e}xico,  University of Arizona, University of Colorado  Boulder, University of Oxford, University of Portsmouth, University of Utah, University of Virginia, University  of Washington,  University of  Wisconsin, Vanderbilt University, and Yale University.
\end{acknowledgments}

\end{document}